\documentclass[epsf,prd,floatfix,nofootinbib]{revtex4}
\usepackage{graphicx}
\usepackage{epsfig}
\usepackage{dcolumn}
\usepackage{bm}
\usepackage{latexsym}

\begin{document}

\title{Limit on the diffuse flux of ultra-high energy
  tau neutrinos with the surface detector of the Pierre Auger Observatory} 

\author{{\bf The Pierre Auger Collaboration} \\
J.~Abraham$^{8}$, 
P.~Abreu$^{71}$, 
M.~Aglietta$^{53}$, 
C.~Aguirre$^{12}$, 
E.J.~Ahn$^{87}$, 
D.~Allard$^{30}$, 
I.~Allekotte$^{1}$, 
J.~Allen$^{90}$, 
P.~Allison$^{92}$, 
J.~Alvarez-Mu\~{n}iz$^{78}$
% ...
}

\affiliation{
~\\
~\\
$^{1}$ Centro At\'{o}mico Bariloche and Instituto Balseiro (CNEA-
UNCuyo-CONICET), San Carlos de Bariloche, Argentina \\
% ...
$^{8}$ Observatorio Meteorologico Parque Gral.\ San Martin (UTN-
FRM/CONICET/CNEA), Mendoza, Argentina \\
% ...
$^{12}$ Universidad Catolica de Bolivia, La Paz, Bolivia \\
% ...
$^{30}$ Laboratoire AstroParticule et Cosmologie (APC), 
Universit\'{e} Paris 7, CNRS-IN2P3, Paris, France \\
% ...
$^{53}$ Istituto di Fisica dello Spazio Interplanetario (INAF),
 Universit\`{a} di Torino and Sezione INFN, Torino, Italy \\
% ...
$^{71}$ LIP and Instituto Superior T\'{e}cnico, Lisboa, Portugal \\
% ...
$^{78}$ Universidad de Santiago de Compostela, Spain \\
% ...
$^{87}$ Fermilab, Batavia, IL, USA \\
% ...
$^{90}$ New York University, New York, NY, USA \\
% ...
$^{92}$ Ohio State University, Columbus, OH, USA \\
% ...
}

\begin{abstract}
Data collected at the Pierre Auger Observatory are used to establish an upper limit on the diffuse flux of tau neutrinos in the cosmic radiation. Earth-skimming $\nu_{\tau}$ may interact in the Earth's crust and produce a  $\tau$ lepton by means of charged-current interactions. The $\tau$ lepton may emerge from the Earth and decay in the atmosphere to produce a nearly horizontal shower with a typical signature, a persistent electromagnetic component even at very large atmospheric depths. The search procedure to select events induced by $\tau$ decays against the 
background of normal showers induced by cosmic rays is described. The method used to compute the exposure for a detector continuously growing with time is detailed. Systematic uncertainties in the exposure from the detector, the analysis and the involved physics are discussed. No $\tau$ neutrino candidates have been found. For neutrinos in the energy range $2\times10^{17}$ eV $< E_{\nu}$ $<$ $2\times10^{19}$ eV, assuming a diffuse spectrum of the form  $E_{\nu}^{-2}$, data collected between 1 January 2004 and 30 April 2008 yield a 90\% confidence-level upper limit of $E_\nu^{2}~\mathrm{d}N_{\nu_\tau}/\mathrm{d}E_{\nu} < 9 \times 10^{-8}$ GeV cm$^{-2}$ s$^{-1}$ sr$^{-1}$.
\end{abstract}

\maketitle

\section{Introduction}

There are many efforts to search for high energy neutrinos with dedicated 
experiments~\cite{Kestel:2004ep,Aggouras:2004mh,Arnold:2004xq,Carr:2006ba,Miocinovic:2005jh}.
Their observation should open a new window 
to the universe since they can give information on regions that are otherwise 
hidden from observation by large amounts of matter in the field of view. Moreover, neutrinos are not deviated by magnetic fields and, hence, they essentially maintain the direction of their production places. The existence of ultra-high energy (UHE) cosmic rays of energies exceeding 
$10^{19}$~eV makes it most reasonable to expect neutrino fluxes 
reaching similar energies. 
Although the origin of cosmic rays and their production 
mechanisms are still unknown~\cite{CRModelsReviews}, neutrinos are expected 
to be produced together with the cosmic rays and also 
in their interactions with the background radiation 
fields during propagation~\cite{NeutrinosReviews}. 
Unfortunately there are still many unknowns 
concerning cosmic ray origin and neutrino fluxes remain quite uncertain. 
Because of their relation to cosmic ray production and transport, 
the detection of UHE neutrinos should in addition give very valuable 
information about cosmic ray origin. 

Models of the origin and propagation of UHECR consider the production
of pions decaying into neutrinos.
If protons or nuclei of extra-galactic origin are accelerated in extreme 
astrophysical environments their interaction with the matter or radiation 
fields in the source region should yield pions which decay giving rise 
to neutrino fluxes. In addition 
cosmic rays interact with the background radiation when traveling 
over long distances 
resulting in a steepening of the spectrum around $5 \times 10^{19}$~eV. 
This is the Greisen-Zatsepin-Kuz'min effect~\cite{Greisen,Zatsepin}, 
consistent with the recently reported suppression of the
cosmic ray flux above $\sim4 \times
10^{19}~$eV~\cite{Abbasi:2007sv,Collaboration:2007ba} as 
well as the observed anisotropy of the highest energy cosmic rays and a possible correlation with 
relatively nearby extragalactic objects~\cite{Collaboration:2007bb,Abraham:2007si}. The GZK mechanism is 
a source of UHE neutrinos, in the case of protons through interactions with  
the Cosmic Microwave Background (CMB) what gives rise to the cosmological 
neutrinos~\cite{Stecker:1991vm} and in the case of iron nuclei through 
interactions with infrared light that dissociates the nuclei. 
Alternative models, often referred to as \emph{top-down} scenarios, have been 
developed  although latest limits on photon fractions~\cite{PAOPhoton} appear to discard them as an adequate explanation of the UHE cosmic rays. They are based on the decay of super-massive particles into leptons and 
quarks. The latter subsequently fragment into cosmic ray protons but pions 
dominate the fragmentation mechanism, their decays giving rise to 
photon and neutrino fluxes. 
The produced neutrinos would exceed those that can 
be expected by the cosmic ray interactions with the background fields and 
are typically produced with harder spectra. 

Both conventional acceleration and top-down scenarios generate pions  
which decay to produce an electron to muon neutrino 
flavor ratio of order 1:2 while neutrinos of $\tau$ flavor are heavily
suppressed at production. With the discovery of neutrino flavor 
oscillations~\cite{Fakuda2001} and maximal $\Theta_{23}$ mixing, the flavor 
balance changes as neutrinos propagate to Earth. After traveling cosmological 
distances approximately equal fluxes for each flavor are 
expected~\cite{Learned:1994wg,Athar2000}. The idea of detecting $\nu_{\tau}$ induced events 
through the emerging $\tau$ produced by neutrinos that enter the Earth just 
below the horizon, was presented for the first time
in~\cite{Letessier-Selvon:2000kk,Fargion:2000iz}. These Earth-skimming 
neutrinos undergo charged-current
interactions to produce a very penetrating $\tau$ lepton. 
When the interaction occurs sufficiently close to the Earth's surface 
the $\tau$ can escape to the atmosphere and decay in flight. This
would in most cases give rise to an extensive air shower traveling
nearly horizontal and in the upward direction for an ideal spherical
Earth's surface.

The Pierre Auger Observatory~\cite{Abraham:2004dt} has been designed to explore ultra high energy 
cosmic rays with unprecedented precision exploiting the two available 
techniques to detect UHE air showers, arrays of particle detectors and 
fluorescence telescopes. It can also detect neutrinos by searching for 
deep inclined showers both with the surface
detector~\cite{Cronin:1999ek} and with fluorescence
telescopes~\cite{Aramo:2004pr}. Showers resulting from $\tau$ decays
induced by Earth-skimming neutrinos can also be detected with the
Pierre Auger Observatory, both with the surface and the fluorescence
detectors. This channel has been shown to increase the possibilities
for detecting neutrinos and in particular using the surface detector
of the Pierre Auger Observatory which becomes most sensitive to
neutrinos in the EeV range~\cite{Bertou2002}. 

An upper limit on the diffuse flux of $\tau$ neutrino of $E_\nu^{2}~\mathrm{d}N_{\nu_\tau}/\mathrm{d}E_{\nu} < 1.3 \times 10^{-7}$ GeV cm$^{-2}$ s$^{-1}$ sr$^{-1}$ at 90 $\%$ C.L. was reported in~\cite{AugerPRLNuTau} using data collected between 1 January 2004 and 31 August 2007 with the surface detector of the Pierre Auger Observatory. The collected data were
searched for $\tau$ neutrino candidates applying a $\nu_{\tau}$
identification criterion that was obtained simulating
Earth-skimming $\nu_{\tau}$s, their interactions in the Earth,
propagation of the subsequent $\tau$ leptons and the associated
showers they produce in the atmosphere. This article discusses in detail the search procedure to discriminate UHE Earth-skimming $\tau$ 
neutrinos used in~\cite{AugerPRLNuTau} as well as the compute of the exposure and the evaluation of the systematics. This article also uses an updated data sample. No candidates have been found
 in data from 1 January 2004 until 30 April 2008 and a new limit to the diffuse flux of
UHE $\nu_{\tau}$ is presented. The article is organized as follows. In
Section \ref{PAO}, the Pierre Auger Observatory is briefly described. In
Section \ref{MCSim}, the needed Monte Carlo simulations are detailed. In
Section \ref{Analysis}, the method for discriminating neutrino-induced
showers is explained and the selection procedure is presented. In 
Section \ref{acceptance}, the computation of the exposure is
reported. In Section \ref{systematic}, the systematic uncertainties are
discussed. In Section \ref{limit}, the results from the Pierre Auger
Observatory data for $\nu_{\tau}$ Earth-skimming neutrinos are
shown. Finally in Section \ref{summary}, this work is summarized.\par

\section{The Pierre Auger Observatory}
\label{PAO}

The Pierre Auger Observatory will consist of two hybrid detectors in the 
northern and southern hemispheres, each one combining an array of particle 
detectors and fluorescence telescopes for redundancy and calibration~\cite{DawsonICRC07}. The Southern Observatory 
is in Malarg\"ue, Mendoza, Argentina and its construction phase is currently completed. It covers 3000~km$^2$ with
regularly spaced particle detectors and with four fluorescence eyes at
the perimeter that overlook the same area~\cite{Abraham:2004dt}. There 
are plans to construct the Northern Auger Observatory
in Lamar, Colorado, USA~\cite{NitzICRC07}. Data have been
taken with the Southern Pierre Auger Observatory since January 2004
while it was under construction. The amount of data that has been
accumulated for the analysis described in this article corresponds
to about $1.5$ times the data that will be gathered in a whole year with the
complete detector. This article will only address the search for
Earth-skimming neutrinos with the array of particle detectors that
constitutes the surface detector of the Southern Pierre Auger Observatory.

\subsection{The Surface Array of the Pierre Auger Observatory}
\label{SD}

The Southern surface detector (SD) consists of 1600 Cherenkov water tanks 
(3.6~m diameter and 1.2~m high) arranged in a triangular grid with 
1.5~km spacing between them, covering an almost flat surface, at an 
approximate altitude of 1400~m above sea level. 
Each tank is a polyethylene tank internally coated with a diffusive 
Tyvek\textsuperscript{\texttrademark} liner filled with 12 tons of 
purified water. The top surface has three Photo Multiplier Tubes (PMTs) 
in optical contact with the water in the tank. The PMT signals are sampled by 
40 MHz Flash Analog Digital Converters (FADC). Each tank is regularly monitored and calibrated in units of Vertical Equivalent Muons (VEM) corresponding to the signal produced by a $\mu$ traversing the tank vertically~\cite{Bertou:2005ze}. The system transmits 
information by conventional radio links to the Central Data Acquisition 
System (CDAS) located in Malarg\"ue. The PMTs, a local 
processor, a GPS receiver and the radio system are powered by 
batteries with solar panels. Once installed, the local stations work
continuously without external intervention. \par

The local trigger at the level of an individual Cherenkov tank (second order 
or T2 trigger) is the logical \emph{OR} of two conditions: either a given 
threshold 
signal (1.75 VEM) is passed in at least one time bin of the FADC trace, or a 
somewhat lower threshold (0.2 VEM) is passed at least in 13 bins within a 3~$\mu$s time 
window (120 bins)~\cite{NitzIEEE}. The latter condition, the so-called \emph{Time over Threshold} (ToT),
is designed to select broad signals in time, characteristic of the early stages of the development of an extensive air shower (EAS). The 
data acquisition system receives the local triggers and builds a global
trigger requesting a relatively compact configuration of 3 local
stations compatible in time, each satisfying the ToT trigger, or 4 
triggered stations with any T2 trigger (a third level or T3
trigger)~\cite{Trigger}. With the complete array, the global T3
trigger rate will be about 3 events per minute, one third being actual
shower events at energies above $3\times10^{17}$ eV.\par

\subsection{The Data Sample}

The SD has been taking data in a stable manner since January 2004~\cite{TinaICRC07}. Meanwhile the array has been growing and the number of deployed stations has increased from 120 to 1600 during the period analyzed in this article.\par

The analysis reported here is restricted to selected periods in order to eliminate  inevitable problems associated to the construction phase, typically in the data acquisition, in the communication system and due to hardware instabilities. To ensure the quality of the data, we have analyzed the arrival time of the events under the reasonable hypothesis that the rate of physics events recorded by the detector (after proper size normalization) is independent of time. Given the large aperture of the SD and the level
at which anisotropies could exist on the sky~\cite{Armengaud:2007hc} this approximation is,
from this point of view, well justified. Assuming a constant rate $\lambda$ of physics events, the probability $P$ of the time interval $t$ between two consecutive events to be larger than $T$ is given by: 

\begin{equation}
\label{eq:BadPeriods}
P(t>T) = e^{-\lambda T}
\end{equation}
where the value of $\lambda$ is the mean rate of the recorded events normalized to the detector size.

Consecutive events for which $P$ is below a certain threshold value $P_{\rm cut}$ are assumed to belong to periods with problems in the data acquisition and are used to define the bad periods to be rejected. The procedure will reject a good period with probability $P_{\rm cut}$ together with the eventual bad ones. So in principle one would like $P_{\rm cut}$ to be as small as possible. The choice of $P_{\rm cut}$ is made by finding where the distribution of probability of the time interval between two consecutive events differs from being flat. The numerical value of $P_{\rm cut}$ was found to be $\sim$10$^{-5}$, which allows us to reject only a small fraction of good periods while removing the periods that lead to a non flat probability. When data taking began, the bad periods were of the order of 10$\%$ of the operating time but by the end of the time period considered in this work we were typically below 1$\%$~\cite{TriggerPaper}.\par

Once the bad periods have been removed, the events that have passed the
third level trigger~\cite{Trigger} from January 2004 until April 2008
constitute the data sample used in this paper.\par

\section{End to end simulation chain}
\label{MCSim}

In order to obtain a flux or a flux limit from the data, the detector neutrino 
showers have to be searched with a selection criterion and the exposure 
of the detector must be accordingly computed. 
Both the criteria to identify neutrino induced showers and the computation
of the exposure to $\nu_{\tau}$ are based on Monte Carlo
techniques. Three separate simulations can be identified. Firstly a dedicated 
simulation that deals with the neutrinos entering the Earth and the  
$\tau$ leptons that exit. A second simulation involves the $\tau$ 
decay in flight and the development of an up-going atmospheric shower. Finally 
a simulation of the tank response to the through-going particles is performed to 
convert the particles at ground level obtained in the shower simulation 
to an actual detector signal. 
\par 

\subsection{Earth-Skimming neutrinos}
\label{EarthSim}

As Earth-skimming $\nu_\tau$s penetrate the Earth they interact to produce 
$\tau$ leptons that suffer energy loss but can escape the Earth 
and decay in the atmosphere. As a result the incoming neutrinos give rise 
to an emerging $\tau$ flux which depends on the depth of matter traversed 
(for a spherical Earth as assumed here, it depends only on the nadir angle). The 
decay of the $\tau$ lepton in the atmosphere induces an extensive air shower 
(EAS) that can trigger the SD. 
The efficiency of this conversion plays a key role in the calculation
of the detector exposure.
The $\tau$ flux has been computed using simulation techniques that 
take into account the coupled interplay between the $\tau$ and the $\nu_\tau$ 
fluxes as they traverse large matter depths through charged current (CC) weak 
interactions and through $\tau$ decay. Energy 
losses induced by neutral current (NC) interactions for both particles are taken into account as a stochastic process.  The energy losses through bremsstrahlung, pair production and nuclear interactions for the $\tau$ lepton are applied continuously, which, at the level of accuracy we need for this work, is a reasonably good approximation~\cite{Bigas:2008ff}. 

%

Propagation of particles through matter is performed in small depth steps. At each step the particles are followed and the probability for interaction 
and decay (in the case of the $\tau$) are evaluated taking into account the 
particle energy. The chain starts with an incident $\nu_{\tau}$ which may
interact by CC or NC. When the former occurs, a $\tau$ lepton is
generated, and its energy is selected taking into account the 
$y$-distribution of the interaction, where $y$ is the fraction of the $\nu_{\tau}$ energy 
transferred to the nucleon in the laboratory frame. 
If the $\nu_{\tau}$ interacts through a NC its energy is computed taking the  
$y$-distribution into account.
Once a $\tau$ lepton is produced, it can undergo energy loss, weak 
interactions both neutral and charged and decay. 
In the case of CC interaction or decay a new $\nu_{\tau}$ is produced 
which regenerates the $\nu_\tau$ flux that is propagated further. Finally, if a
$\tau$ lepton emerges from Earth, its energy, direction and decay 
position are stored and used as an input for the simulation of
atmospheric showers induced by $\tau$ leptons.\par

For the relevant depths inside the Earth where $\tau$ leptons can be produced and reach the surface (few km) a homogeneous density of 2.65 g cm$^{-3}$ can be assumed. Parameterisations of 
the cross section for weak interactions and for the $y$-distributions at 
very high energy are used. 
The cross section for CC interactions is taken from~\cite{CooperSarkar:2007cv} 
and the $y$-distribution from~\cite{Gazizov:2004va}. For NC
interactions, the cross section is assumed to be 0.4 that of the
CC~\cite{Gandhi98}. The energy losses for $\tau$ leptons are
parameterised following case III in~\cite{Dutta:2005yt}, which gives
the best representation of Monte Carlo simulation.\par 

\subsection{Extensive Air Showers in the atmosphere}
\label{SimEAS}
The $\tau$ decays in the atmosphere give rise to secondaries that may 
initiate an EAS that can trigger the SD. 
The decay mode has been simulated using the TAUOLA package version 
2.4~\cite{TAUOLA} to obtain the type of the secondaries and their energies  
which are subsequently injected in Aires (version 2.6.0)~\cite{Aires260} with SIBYLL 2.1~\cite{sibyll} as a model for the hadronic interactions at high energy. 
Showers induced by up-going $\tau$s with energies from 
log(E$_{\tau}$/eV) = 17 to 20.5 in steps of 0.5 have been simulated 
at zenith angles ranging between $90.1^\circ$ and $95.9^\circ$ 
in steps of $0.01$ rad and at an altitude above the Pierre Auger 
Observatory that ranges from $0$ m to $2500$ m in $100$ m steps. Ten
showers have been generated for each combination of energy, zenith
angle and altitude, which leads to a total of 20 000 showers.

The extremely large amount of particles involved in an $\sim$ EeV shower 
makes it impractical to follow all the secondaries. 
The current simulation packages include a
statistical sampling algorithm based on the thinning algorithm originally
introduced in~\cite{Hillas1981}. Only a small representative fraction
of the total number of particles is propagated. Statistical 
weights are assigned to sampled particles in order to compensate for
the rejected ones. \par

\subsection{Detector response}

The first step in the detector simulation is to obtain the particles
reaching each tank from the sampled particles produced in the
simulation of the EAS. A re-sampling algorithm is 
necessary to convert the output of the program to the expected number of 
particles that enter a SD station. 
This is done averaging over an area around the station that is
large enough to avoid unphysical fluctuations from the thinning
procedure, and at the same time small enough to avoid large
differences in the density and average properties of particles in different places on the
area~\cite{billoir2008}. Each particle reaching the station is injected inside the tank,
and a detailed simulation is performed to obtain the light hitting 
the PMTs as a function of time. The simulated FADC traces are obtained as 
the superposition of the signal of all individual particles entering the tank 
accounting for their arrival time. Finally  
both the local and central trigger algorithms are applied and the 
event is stored in the same format as data~\cite{GhiaICRC07}.\par

At the highest simulated values of incident angles or altitudes where
the $\tau$ decays, none of the simulated showers at any of the 
simulated energies fulfills the central trigger 
conditions. This is taken as a clear indication that a 
complete sample of showers has been produced without introducing any bias 
and that it therefore correctly represents the 
characteristic $\tau$ showers that could trigger the SD 
detector.\par

\section{Discrimination of neutrino-induced showers}
\label{Analysis}

\subsection{Neutrino signature: inclined showers in the early stages}
\label{signature}

\begin{figure}[h]
  \centering
  \includegraphics[width=0.8\textwidth]{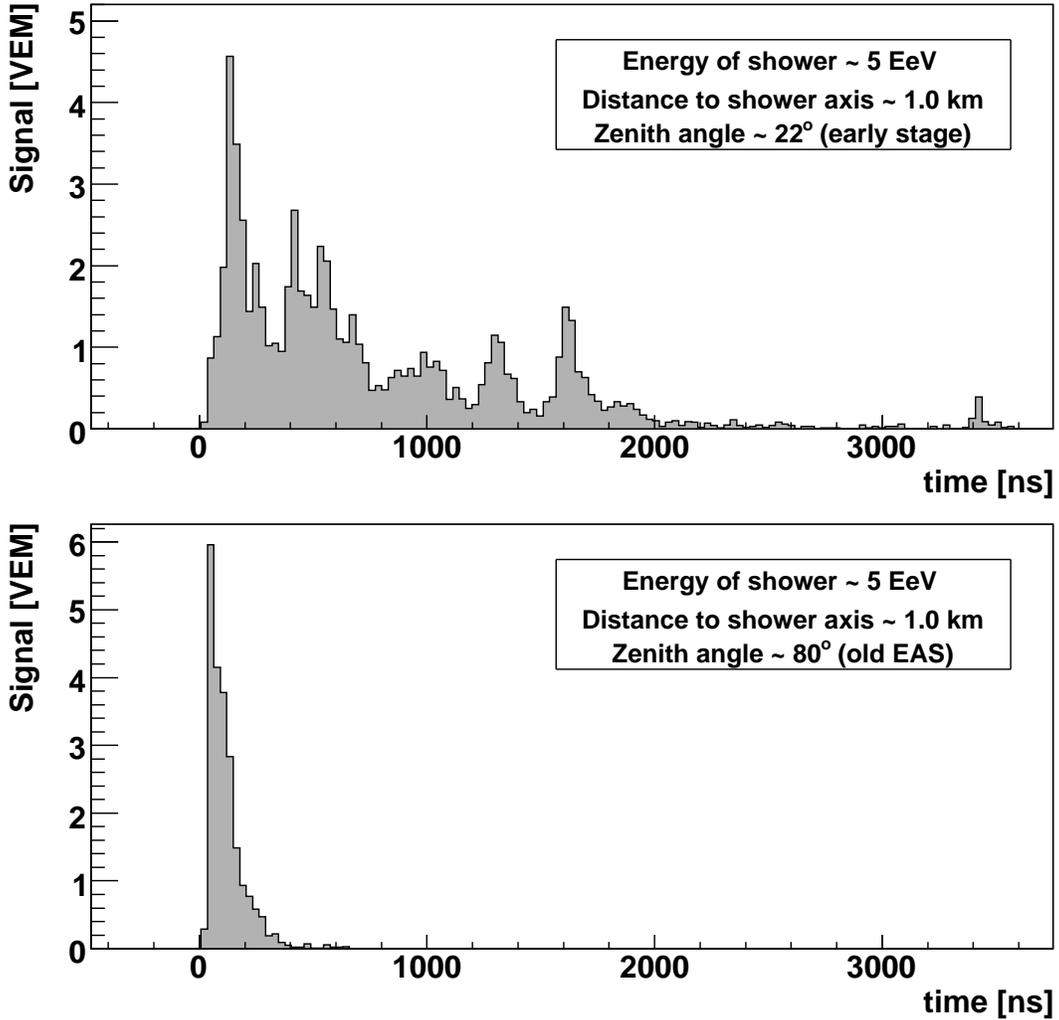}
  \caption{FADC traces of stations at 1 km from the shower
core for two real showers of 5~EeV. Top panel: early stages of development ($\theta\sim$ 22$^{\circ}$); bottom: old extensive air shower ($\theta\sim$ 80$^{\circ}$).}
   \label{traces}
\end{figure}

UHE particles interacting in the atmosphere give rise to a shower
with an electromagnetic component reaching its maximal development
after a depth of the order of 800 g cm$^{-2}$ and extinguishing
gradually within the next 1000 g cm$^{-2}$. After roughly a couple 
of vertical atmospheric depths only high energy muons survive. In the
first stages of development, while the electromagnetic component
develops, the time spread of the particles in the shower front is
large ($\sim \mu s$). When the shower becomes old, most of the particles in the shower front, the high energy muons, arrive in a short time window ( $\sim$ 100 ns ).
As a consequence very inclined showers induced by protons or nuclei 
(or possibly photons) in the upper atmosphere reach the ground as a
thin and flat front of muons accompanied by an electromagnetic halo, which is produced by bremsstrahlung, pair production and muon decays, and has a time structure very similar to that of the muons. On the other hand, if a shower is induced by a
particle that interacts deep in the atmosphere (a deep neutrino interaction
in air, or a tau decay), its electromagnetic component could hit the
ground and give a distinct broad signal in time. The signal in each station of
the SD is digitized using FADCs, allowing us
to unambiguously distinguish the narrow signals from the broad ones
and thus to discriminate stations hit by an EAS in the early stages of development or by an old EAS. This is illustrated in figure \ref{traces} where we show 
FADC traces from two different real events. The FADC trace taken from the shower with a zenith angle  of 22$^\circ$ is representative of an EAS in the early stages while the other is representative of an old EAS.\par

\subsection{Identification of neutrino candidates}
\label{Identification}
The identification of showers induced by Earth-skimming $\tau$ neutrinos implies
searching for very inclined (quasi-horizontal) showers in an early stage of development. Broad signals, which are characteristic as long as the electromagnetic component still develops, produce a ToT local trigger (see section \ref{SD}). A bunch of  muons from cosmic ray showers can produce high amplitude
signals extended in time or two independent muons can arrive inside the 
given time interval. Both would also produce a ToT local trigger which 
is not associated to the presence of electromagnetic component from a 
neutrino shower. To get rid of them, a further requirement is made to
the signals in order to filter out these backgrounds. First a cleaning of the FADC trace is done to remove segments of the trace that could be generated by an accidental muon arriving closely before or after the shower front. Segments of the FADC trace are defined by neighbour bins above 0.2 VEM, allowing gaps of up to 20 bins and only the segment with the largest signal is kept. An \emph{offline ToT} is defined by requiring that the signal after cleaning (the segment with largest signal) of the FADC trace has at least 13
bins above the low threshold (0.2 VEM) and the ratio of the
integrated signal over the peak height exceeds by a factor 1.4 the average ratio observed in
signals of isolated particles (as defined in the calibration 
procedure~\cite{Bertou:2005ze}). The central trigger conditions are
applied only to stations that fulfill the offline ToT. Still a small number of 
nucleonic showers with a large number of triggered tanks may have a subsample 
of stations that satisfy this condition even if in all the other stations the 
signal is not broad at all. In order to reject such events, at least
60$\%$ of the triggered stations are required to fulfill the offline
ToT. After this selection procedure, an almost pure sample of showers reaching the ground at their early stages is selected. \par 

\begin{figure}[h]
   \centering
  \includegraphics[width=0.8\textwidth]{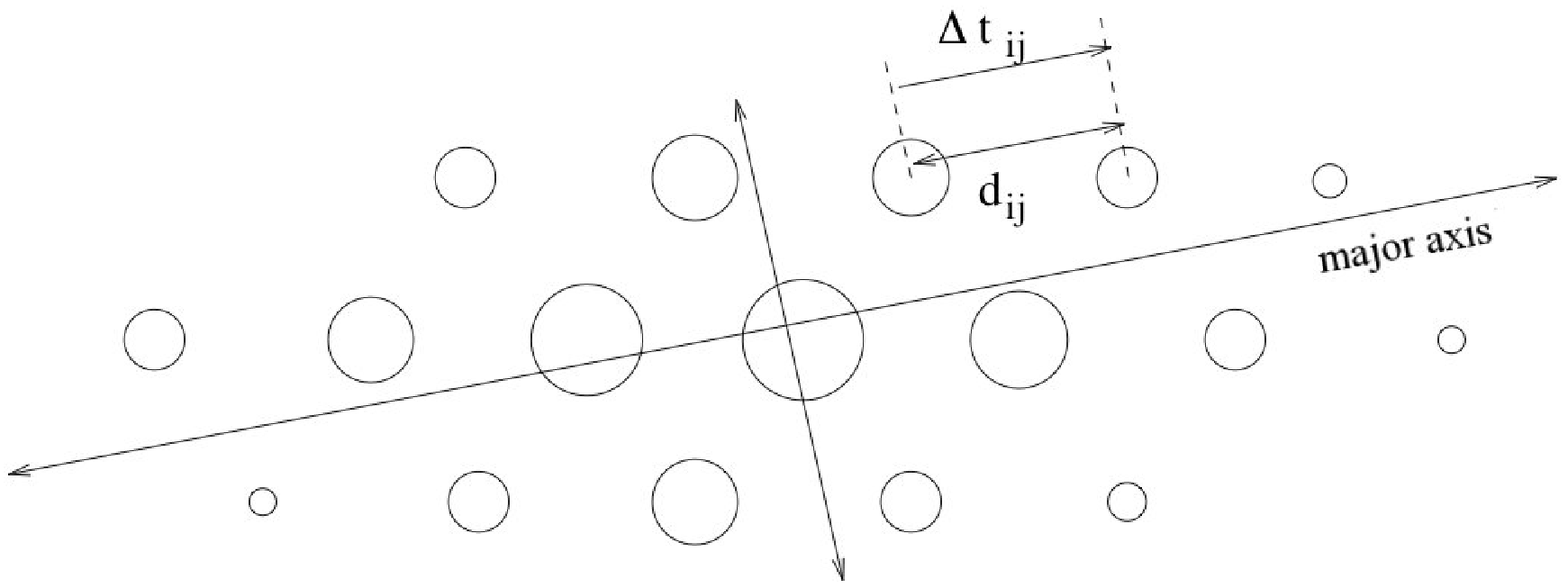}
  \caption{Schematic view of the footprint of a shower on the SD
  array. Each circle represents the position of a station, and their
  sizes are proportional to the station signal.}
  \label{moments}
\end{figure}

Once the criterion for young showers is established a second criterion must 
be used to select very inclined showers as expected from Earth-skimming 
neutrino interactions. The devised method uses two variables associated 
to the \emph{footprint} that the triggered tanks of the event leave on the 
ground and the apparent speed with which the signal \emph{moves} across the 
array. Firstly a symmetric tensor is 
built using the station signals included in the central trigger 
and their ground positions (analogous to the tensor of inertia of a 
mass distribution, see Eq.~(\ref{Eq:Tensor})). 

\begin{eqnarray}
S = \Sigma_{i} s_{i} \hspace{1cm} \langle X \rangle = \Sigma_{i} s_{i} x_{i} / S  \hspace{1cm} \langle Y \rangle = \Sigma_{i} s_{i} y_{i} / S  
\nonumber \\
 I_{xx} = \Sigma_{i} s_{i} (x_{i}-\langle X \rangle)^{2}/S \hspace{1cm} I_{yy} = \Sigma_{i} s_{i} (y_{i}-\langle Y \rangle)^{2}/S
\nonumber \\
 I_{xy} = I_{yx} = \Sigma_{i}^{n} s_{i} (x_{i}-\langle X \rangle)(y_{i}-\langle Y \rangle) \hspace{1.5cm}
\label{Eq:Tensor}
\end{eqnarray}
where $s_{i}$ is the signal in VEM for each station; ($x_{i},y_{i}$) are the coordinates of each station; and $\Sigma_{i}$ is the sum over the stations.

The corresponding major and minor axes are used to define a characteristic \emph{length} and a \emph{width} of the pattern as the square root of the eigenvalues of the symmetric tensor (see Eq.~(\ref{Eq:LW})).
\begin{eqnarray}
\label{Eq:LW}
\nonumber \\
length^{2} = \frac{I_{xx}+I_{yy} + \sqrt{( I_{xx}-I_{yy})^2+4I_{xy}^{2}}}{2S} 
\nonumber \\
width^{2} = \frac{I_{xx}+I_{yy} - \sqrt{( I_{xx}-I_{yy})^2+4I_{xy}^{2}}}{2S} 
\end{eqnarray}

Secondly for each pair of tanks $(i,j)$, a \emph{ground speed} is defined as $d_{i,j}/|\Delta t_{i,j}|$, where $d_{i,j}$ is the distance between the tanks 
(projected onto the major axis) and $|\Delta t_{i,j}|$ is the difference 
between the start times of their signals (figure \ref{moments}). 
Quasi-horizontal showers have an elongated shape 
(characterized by a large value of  \emph{length}/ \emph{width}) and 
they have ground speeds tightly concentrated around the speed of 
light c. 

\begin{figure}[h]
   \centering
  \includegraphics[width=0.8\textwidth]{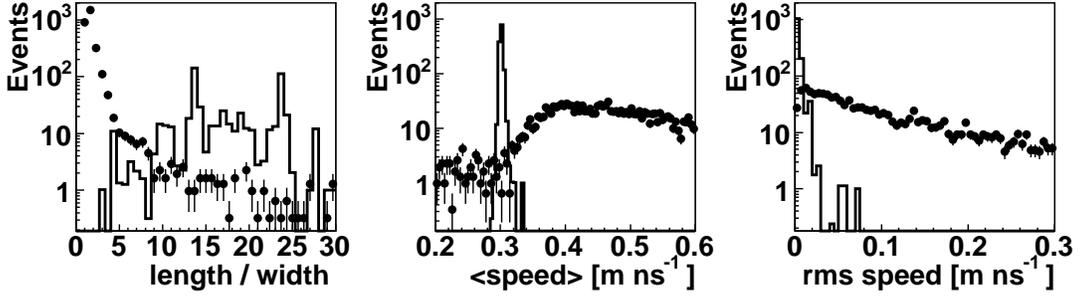}
  \caption{Distribution of variables used to discriminate very inclined showers for an incident  
  $E^{-2}$ $\nu_\tau$ flux (histogram), and for real events collected during November and December 2004 passing the
  \emph{early stage} (see text) selection (points). Left panel: length/width;
  middle: average of the ground speed between pairs of stations; right:
  r.m.s. of the ground speeds.}
  \label{variables}
\end{figure}
In figure \ref{variables} the distributions of the two 
discriminating variables are shown for real events and simulated tau
showers. Based on the comparison between MC simulations and data collected during November and December 2004, which is less than 1$\%$ of the used data sample, the following cuts were fixed to select Earth-skimming tau neutrino candidates: 
\begin{itemize}
\item{} \emph{length}/\emph{width} $>$ 5
\item{} 0.29 m ns$^{-1}$ $<$ average \emph{ground speed} $<$ 0.31 m ns$^{-1}$
\item{} r.m.s. (\emph{ground speed}) $<$ 0.08 m ns$^{-1}$
\end{itemize}
where the average \emph{ground speed} and its dispersion are computed using only stations for which $|d_{i,j}|$ is larger than 1000 m.

Since the selection criteria relies on the shower footprint, we need to guarantee that a representative fraction of the event is detected with the
SD. For this purpose the closest station to the center of the footprint 
(values  $\langle X \rangle$ and  $\langle Y \rangle$ defined in Eq.~\ref{Eq:Tensor})
is required to be 
surrounded by at least five working stations at the time of
occurrence of the event. Hence, events at the edges of the array 
with small detected fraction of the footprint typically do not fulfill 
the selection criteria. 
This procedure is simple and robust. It can be applied to any footprint
and does not require any global reconstruction. \par

\section{Neutrino Exposure of the Surface Detector of the Pierre Auger Observatory}
\label{acceptance}

\begin{figure}[th]
  \begin{center}
    \includegraphics[width= 0.8\textwidth]{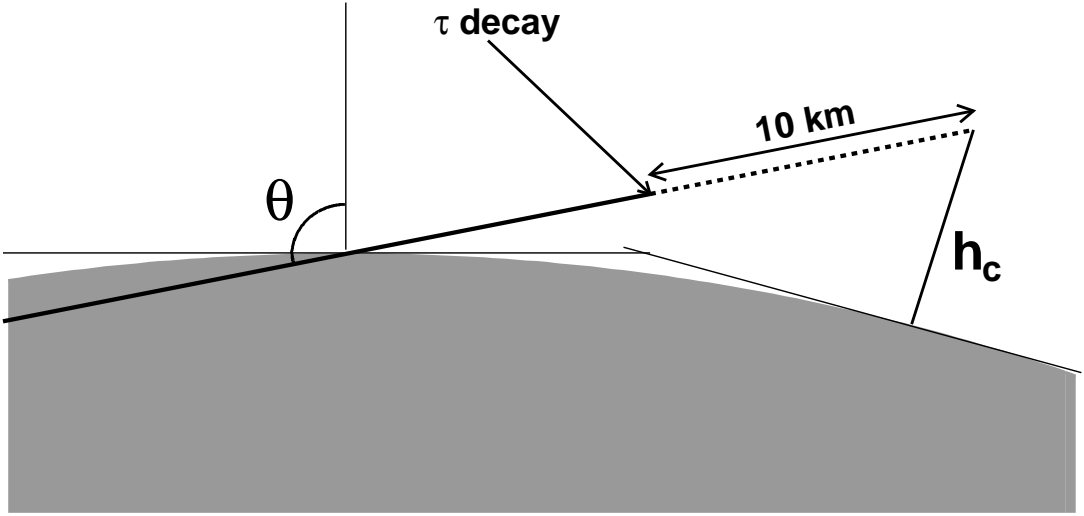}
  \end{center}
  \vspace{-0.5pc}
  \caption{Geometry of the induced $\tau$ shower with the definition of the parameters, $h_{c}$ and $\Theta$, involved on the exposure computation (see text). Angles and distances are not proportional. They have been exaggerated to help the readability of the figure.}
\label{fig:sketch}
\end{figure}

The next step in the calculation is to compute the exposure of the
SD of the Pierre Auger Observatory to showers induced by 
UHE $\nu_{\tau}$. Each simulated Earth-skimming  $\nu_{\tau}$ event
has to be tracked from the 
injection up to its identification through the defined selection cuts. 
The number of identified events is computed from the simulations of the EAS 
initiated by the secondaries in the tau decay, and from the detector 
response to them. 
For a fixed energy of the $\tau$ ($E_{\tau}$),
there is effectively only one relevant parameter determining the efficiency 
of trigger and identification: the altitude of the shower center ($h_{c}$). 
This is conveniently defined as the altitude of the shower axis at a distance of 10 km away from the
$\tau$ decay point along the shower axis (see figure~\ref{fig:sketch}). For  
the shower energies relevant in this analysis, $h_{c}$ is very close to the 
altitude at which a horizontal shower has the largest lateral extension and 
is thus capable of producing the largest footprint at
ground~\cite{Bertou2002}.\par 

\begin{figure}[th]
  \begin{center}
    \includegraphics[width= 0.8\textwidth]{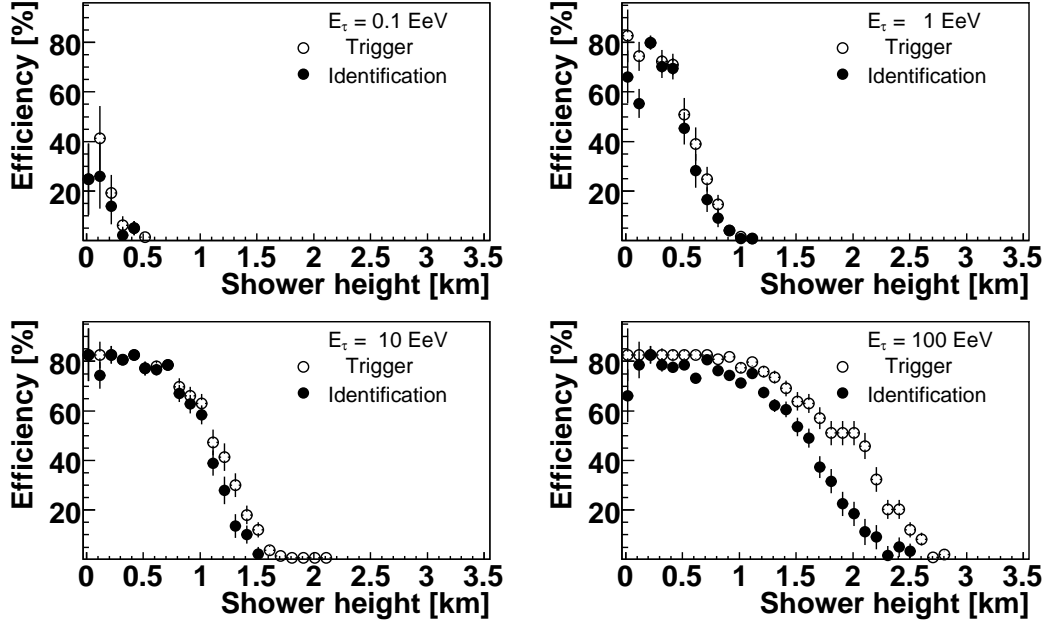}
  \end{center}
  \vspace{-0.5pc}
  \caption{Trigger (open dots) and identification (closed dots)
  efficiency as a function of the height above ground of the shower at
  10 km from the decay point. The efficiency is shown for MC showers induced by
  $\tau$s with energy of 0.1 (top-left), 1 (top-right), 10
  (bottom-left) and 100 (bottom-right) EeV merging all zenith angles.}
\label{fig:TrigEff}
\end{figure}

For a SD that covers a surface $A$, the aperture for a given neutrino
energy ($E_{\nu}$) can be expressed as follow:

\begin{eqnarray}
\label{Aperture1}
Ap(E_{\nu})  =  \int A cos\Theta d\Omega \int_{0}^{E_{\nu}}\hspace{-0.2cm}dE_{\tau}\int_{0}^{\infty}\hspace{-0.2cm}dh_{c}
  \left(\frac{d^2p_{\tau}(E_{\nu})}{dE_{\tau}dh_{c}}\right) \epsilon_{ff}\hspace{-0.8cm}
 \nonumber \\
   =  2 \pi \int_{\pi / 2+\alpha_{m}}^{\pi / 2}\hspace{-0.8cm}A  cos\Theta sin\Theta d\Theta \int_{0}^{E_{\nu}}\hspace{-0.2cm}dE_{\tau}\int_{0}^{\infty}\hspace{-0.2cm}dh_{c}
  \left(\frac{d^2p_{\tau}(E_{\nu})}{dE_{\tau}dh_{c}}\right) \epsilon_{ff} \hspace{-0.8cm}
  \nonumber \\
  =  \pi A sin^{2}\alpha_{m} \int_{0}^{E_{\nu}}\hspace{-0.2cm}dE_{\tau}\int_{0}^{\infty}\hspace{-0.2cm}dh_{c}
  \left(\frac{d^2p_{\tau}(E_{\nu})}{dE_{\tau}dh_{c}}\right) \epsilon_{ff} \hspace{1.0cm}
\end{eqnarray}
where $d^2p_{\tau}/(dE_{\tau}dh_{c})$ is the differential probability of an 
emerging $\tau$ as a function of energy and altitude for a fixed incident 
$\nu_\tau$ energy, that can be easily obtained folding the simulations 
described in section~\ref{EarthSim} for the emerging $\tau$s with the tau decay probability as 
a function of flight distance. $\epsilon_{ff}$ is the probability to
identify a $\tau$ (including the trigger efficiency), that is assumed to 
depend only on $E_{\tau}$ and $h_{c}$. The integral in $\Theta$ is
done from $\pi$/2+$\alpha_{m}$ rad ($\alpha_{m}$=$0.1$ rad) to
$\pi$/2 rad since an incident $\nu_{\tau}$ with a greater angle has
no chance to produce an emerging $\tau$ that produces an observable
shower at ground level. The latter integration can be performed by the
Monte Carlo technique as described in section \ref{MCSim}, leading to: 

\begin{equation}
\label{Aperture1bis}
Ap(E_{\nu}) = \pi A \sin^{2}\alpha_{m}\frac{\sum_{i}
  \epsilon_{ff}(E_{\tau}^{(i)},h_{c}^{(i)})}{N_{sim}}
\end{equation}
where $N_{sim}$ is the number of simulated events. In figure \ref{fig:TrigEff},
the trigger and identification efficiencies for an ideal (no holes and no malfunctioning stations) and
infinite array are shown as given by the simulated EAS described in section
\ref{SimEAS}. They have been calculated by throwing once each simulated EAS on the detector array with a random core position.  The maximum efficiency that can be reached is 82.6 $\%$ due to the $\mu$ channel
decay~\cite{Yao:2006px}. This decay mode does not produce a detectable shower
neglecting the possibility of hard muon bremsstrahlung or pair
production near the detector which should have a negligible effect
on the final limit. The identification efficiency depends smoothly
on $E_{\tau}$ and $h_{c}$, and hence it can be safely
interpolated.\par 

During the period of data taking considered in this work, the SD of the 
Pierre Auger Observatory has been growing continuously. It is of course 
mandatory to take into account this evolution, as well as the 
instabilities of each station. Therefore Eq. 
(\ref{Aperture1bis}) is not valid to compute the actual SD array. 
Instead the following expression can be used:\par

\begin{eqnarray}
\label{Aperture2}
Ap(E_{\nu},t) = \hspace{6.3cm} \\
= \hspace{-0.1cm}\pi \sin^{2}\hspace{-0.1cm}\alpha_{m}
 \hspace{-0.15cm}\int_{0}^{E_{\nu}}\hspace{-0.5cm}dE_{\tau}\hspace{-0.15cm}\int_{0}^{\infty}\hspace{-0.45cm}dh_{c}
\hspace{-0.1cm} \left(\hspace{-0.15cm}\frac{d^2p_{\tau}}{dE_{\tau}dh_{c}} \hspace{-0.1cm} \int_{S} \hspace{-0.15cm}  
 dx dy \, \epsilon_{ff}(E_\tau\hspace{-0.05cm},\hspace{-0.05cm}h_c\hspace{-0.05cm},\hspace{-0.05cm}x\hspace{-0.05cm},\hspace{-0.05cm}y\hspace{-0.05cm},\hspace{-0.05cm}A_{Conf}(t))\hspace{-0.15cm}\right) \hspace{-1.0cm}\nonumber 
\end{eqnarray}
where $\epsilon_{ff}$ now also depends on the position of the shower in the
array $(x,y)$, and on the instantaneous configuration of the array at time
$t$ denoted here as $A_{Conf}(t)$. The integral over the area 
$S$ includes the whole SD array. 

Hence, the total exposure during the considered period of data taking
is the time integration of the instantaneous aperture given by: 

\begin{eqnarray}
\label{Acceptance}
Exp=\hspace{-0.1cm}\int \hspace{-0.1cm} dt \, Ap(E_{\nu},t) \nonumber = \hspace{-0.1cm}\pi \sin^{2}\alpha_{m}
\hspace{-0.1cm} \int_{0}^{E_{\nu}}\hspace{-0.3cm}dE_{\tau}\int_{0}^{\infty}\hspace{-0.3cm}dh_{c}
 \left(\frac{d^2p_{\tau}}{dE_{\tau}dh_{c}} \bar{B}_{\tau}\right) \\
 \bar{B}_{\tau}(E_{\tau},h_{c}) = \hspace{-0.2cm} \int_{T}\hspace{-0.1cm} dt 
 \int_{S} dx dy \, \epsilon_{ff}(E_{\tau},h_{c},x,y,A_{Conf}(t))\hspace{0.8cm}
\end{eqnarray}

\begin{figure}[th]
  \begin{center}
    \includegraphics[width= 0.8\textwidth]{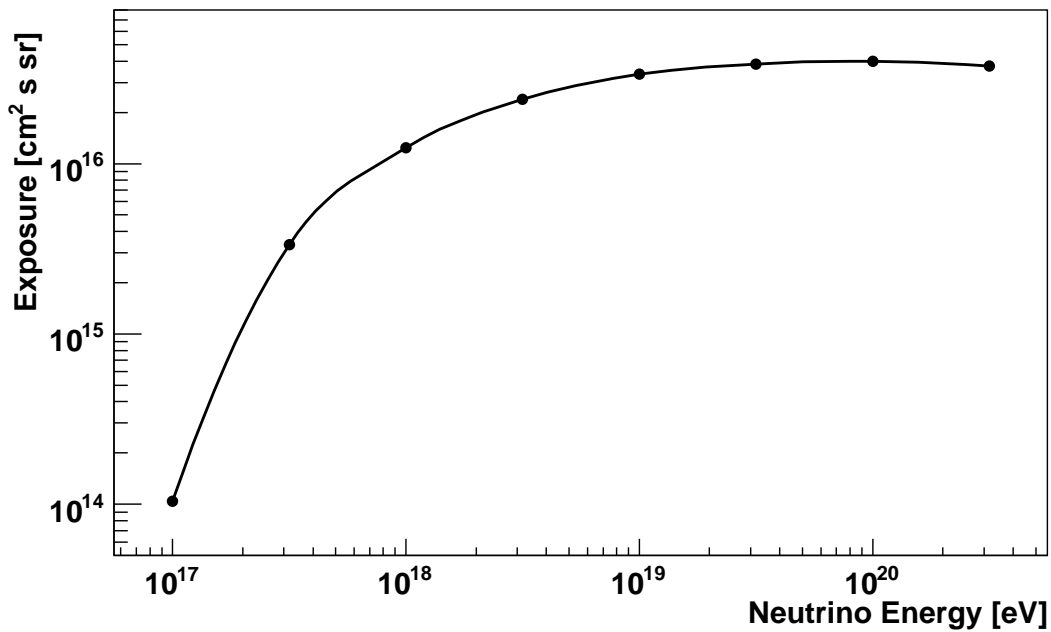}
  \end{center}
  \vspace{-0.5pc}
  \caption{Earth-Skimming neutrino exposure of the Pierre Auger
  Observatory accumulated from January 2004 until April 2008. All the
  identification cuts described in setion \ref{Identification} are
  taken into account.}
\label{fig:Acc}
\end{figure}

The exposure is computed by Monte Carlo in two independent
steps. First, the integrals in $t$ and $(x,y)$ are computed using the
simulations of the EAS and the detector. The number of working 
stations and their status are monitored every second allowing us to
know with very good accuracy the instantaneous SD configuration~\cite{Aperture}. 
For each simulated EAS, several random times from
January 2004 until April 2008 excluding the rejected periods are selected. The number of random times is selected in a monthly base to ensure a statistical precision on the exposure at 1$\%$ level. For each time, the
evaluation of the identification efficiency is done for any
position of the shower in the SD array. The average over all showers
with the same $E_{\tau}$ and $h_{c}$ gives the integral in time and area of  
$\epsilon_{ff}$, allowing one to compute $\bar{B}_{\tau}(E_{\tau},h_{c})$. The
second step computes the integral in $h_{c}$ and $E_{\tau}$ as in 
the case of a perfect array. The estimated uncertainty of this method given the Monte Carlo simulations is below 3$\%$. The accumulated exposure is shown in 
figure \ref{fig:Acc}. It corresponds to an equivalent time of about $1.5$ years of
the complete SD array (1600 water tanks).\par 

\section{Systematic uncertainties}
\label{systematic}

Several sources of systematic uncertainty have been carefully considered. They are addressed below. We have chosen as a reference the aperture calculated with $\nu$ cross section from Ref.~\cite{CooperSarkar:2007cv}, the parameterization of the energy losses from Ref.~\cite{Dutta:2005yt}, an uniform random distribution for the $\tau$ polarization, a spherical model of the Earth and the SIBYLL~\cite{sibyll} hadronic model in combination with AIRES shower simulator~\cite{Aires260}. The systematic uncertainties in this section are all quoted with respect to this aperture and therefore in general asymmetric. Moreover, to be able to quote a range for the systematic uncertainties independently of the energy, an $E^{-2}$ incident flux of neutrinos has been assumed.

Firstly, the location of the Pierre Auger Observatory is 
close to the Andes and not very far away from the
Pacific. The actual topography of the Pierre Auger Observatory can be
taken into account by a detailed Monte Carlo
simulation~\cite{Gora:2007nh}. The effect of the Andes on the expected
event rate has been studied with the aid of a digital elevation map available 
from
the Consortium for Spatial Information~\cite{CGIARCSI}. The number of
detected events decreases by 18 $\%$ if Andes are neglected. \par

There is quite some level of uncertainty in EAS simulation because
accelerator data have to be extrapolated to the shower energies under
discussion. However these uncertainties are not expected to 
have a large effect on the final result since the electromagnetic component 
of the shower, which is the most relevant part for neutrino identification, 
is believed to be better reproduced by simulations. 
Shower simulations have been done with two
hadronic models (QGSJET~\cite{qgsjet} and SIBYLL~\cite{sibyll}) and
passed through two different detector simulations. Based on that, 
systematic uncertainties of 
${+20\%}\atop{-5\%}$ are quoted as due to the 
Monte Carlo simulation of both the EAS and the detector, the
former being the main contribution\footnote{Currently, QGSJET and SIBYLL are the only hadronic models available to be used with the used EAS simulation package. Other recently introduced models like EPOS~\cite{Werner:2007vd} are not yet available to test their effect. Although, in the case of EPOS, due to the large number of muons, and the flatter lateral distribution the trigger efficiency should be larger and in this respect our limit should be conservative.}. 
The simulations of the interactions inside the Earth have been extensively 
checked by comparison with an analytical calculation~\cite{Zas}, an iterative 
solution of the transport equations~\cite{Bigas:2008sw}, and several independent 
simulations. The uncertainty associated to the simulation process itself 
is expected to be below the 5 $\%$ level.\par

Monte Carlo simulations also make use of several physical
magnitudes that have not been experimentally measured at the relevant energy
range, namely: the $\nu$ cross-section, the $\tau$ energy losses and
the polarisation of the $\tau$. All of them can be computed in the
framework of the Standard Model of particle physics using the parton
distribution functions (PDFs).\par

The allowed range for the $\nu$ cross section due to uncertainties in the PDFs has been studied 
in~\cite{CooperSarkar:2007cv} and includes both the effects of the
experimental uncertainties on the PDFs fitted to ZEUS and fixed target
data evolved at next-to-leading-order~\cite{Chenakov03}, as well as theoretical
uncertainties in the implementation of heavy quark masses on the PDF
evolution. For the purpose of this study, the ZEUS PDFs and their uncertainties
were recalculated ~\cite{CooperSarkar:2007cv} using the DGLAP equations throughout the
relevant kinematic range (down to $x \sim 10^{-12}$). This leads to a
$^{+5\%}_{-9\%}$ systematic uncertainty for the number of $\nu$
expected to be detected by the SD of the Pierre Auger Observatory.\par

The decay of the $\tau$ lepton plays a key role on the whole Monte
Carlo simulation. Both the branching ratios of the different decay
modes and the energy distribution among the products are
important. The latter depends on the $\tau$ polarisation, which in turn depends
on the PDFs. The most and least favorable cases in the range of possible 
polarisations (helicity $\pm$1) have been used to estimate the uncertainty
associated to it. The use of the extreme cases of polarisation of the
$\tau$ will not produce more than 
${+17\%}\atop{-10\%}$ 
differences on the exposure.\par

Finally, energy losses include $\tau$ bremsstrahlung (BS)
and pair production (PP) as well as nuclear interactions. 
The contributions from BS and PP can be accurately 
rescaled from the values for muons~\cite{Yao:2006px,VanGinneken:1986rf}. 
The nuclear contribution comes from the photo-nuclear 
cross-section and it is much more uncertain. The differential photo-nuclear 
cross-section as a function of the PDFs has been given
in~\cite{Dutta:2000hh,Bugaev:2002gy}. There exist estimates of the
tau energy losses for the relevant energy range based on 
them~\cite{Aramo:2004pr,Dutta:2005yt,Dutta:2000hh,Bugaev:2002gy,Bugaev:2003sw}.
Different calculations of the energy losses may lead up to 
${+25\%}\atop{-10\%}$ 
systematic uncertainties in the exposure.\par

\begin{figure}[ht]
  \centerline{\includegraphics[width=0.8\textwidth]{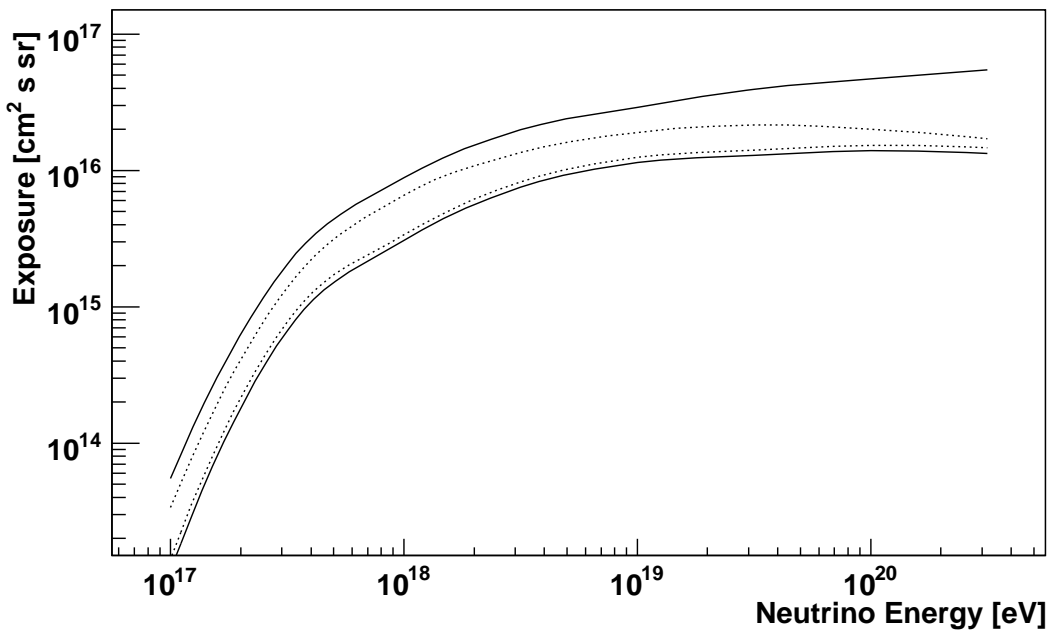}}
  \caption{The Pierre Auger Observatory exposure for one year of a
      full SD array data. Solid lines bracket the total systematic
      uncertainty, while the dotted ones only bracket the allowed range of polarisations,
      $\nu$ cross-sections and energy loss uncertainties.}
  \label{fig:SysAcc}
\end{figure}

In figure \ref{fig:SysAcc} the exposure for 1 year of the SD array 
with 1600 water tanks is shown in the most and least favorable cases of 
the systematic uncertainties previously discussed. 
The systematic uncertainties do not have the same effect for all $\nu_\tau$
energies. The importance of each different contribution to the global systematic uncertainty in the exposure is neither the same at all $\nu_\tau$ energies. At low energies ($\sim$1 EeV) the $\tau$ polarization, the $\nu$ cross section and the $\tau$ energy losses dominate. At higher energies those contributions become smaller and other increase. The latter comes from neglecting the mountains, the effect of which increases with the $\nu_\tau$ energy. The small depth traversed becomes more relevant due to the larger cross section. The former is due to the contribution from the $\nu$ cross section uncertainties. A larger cross section increases the interaction probability for the neutrinos but reduces the solid angle due to the flux absorption in the Earth. This makes the uncertainty in the cross section to contribute mainly around 1~EeV.

The effect of the systematic uncertainties on the expected rate of identified $\nu_\tau$ will depend on the shape of the
actual incident $\nu_\tau$ flux. The effect is almost the same either for GZK-like fluxes or  for
$E^{-2}$ fluxes,  giving a factor $\sim 3$ for the systematic uncertainty in either case (see Table \ref{tab:summary}). Moreover, the energy dependent effect also produces differences on the energy range where most of the identified $\nu_{\tau}$ are expected. For instance, if an $E^{-2}$ flux is assumed the energy range where 90$\%$ of the events are expected changes from 0.22-23 EeV in the least favorable scenario to 0.20-26 EeV in the most favorable one. \par

\begin{table}[h]
\begin{center}
\fontsize{9}{10} \selectfont
\begin{tabular}{|l|c|}
  \hline
  Source & factor \\
  \hline 
  EAS Simulations & 1.30\\
  Topography & 1.18 \\
  Tau Polarisation & 1.30 \\
  Cross Section & 1.15 \\
  Energy losses & 1.40 \\
  \hline
  Total & 2.9 \\
  \hline
\end{tabular}
\caption{Ratio of expected number of $\nu_{\tau}$ 
  for either GZK-like or for $E^{-2}$ incident spectra in the most and least
  favorable scenarios for each source of systematic
  uncertainties.}
\label{tab:summary}
\end{center}
\end{table}

The relevant range of PDFs involved in both the $\nu_\tau$ and the
$\tau$ photo-nuclear cross-sections includes combinations of Bjorken-$x$ and 
$Q^{2}$ where no experimental data is available. Only extrapolations that follow the behavior observed in the regions with experimental data have been considered. Different
extrapolations to low $x$ and high $Q^{2}$ would lead to a wide range
of values for the $\nu$ cross-section as well as for the $\tau$ energy
losses. The systematic uncertainties due to this  have not been included 
in the quoted systematics. Possible large $\nu$
cross-sections have not been taken into account either.


\section{Search results and neutrino limit}
\label{limit}

The measurement of the spectral shape or energy dependent upper limits
on $\nu_\tau$ are out of the reach of Pierre Auger Observatory scope 
for mainly three reasons. Firstly the generic energy reconstruction algorithms
developed for conventional nucleonic showers do not work if the
position of the shower axis is not determined (here, a 
knowledge of the altitude at which the shower is produced would be needed); 
secondly, the fraction of $\tau$ energy contributing to the EAS depends on the
decay mode; and finally, the energy transferred from the incident $\nu_{\tau}$ 
to the emerging $\tau$ is not known. At best an approximate lower bound of 
the initial $\nu_\tau$ energy could be obtained.\par

\begin{figure}[h]
   \centering
  \includegraphics[width=0.8\textwidth]{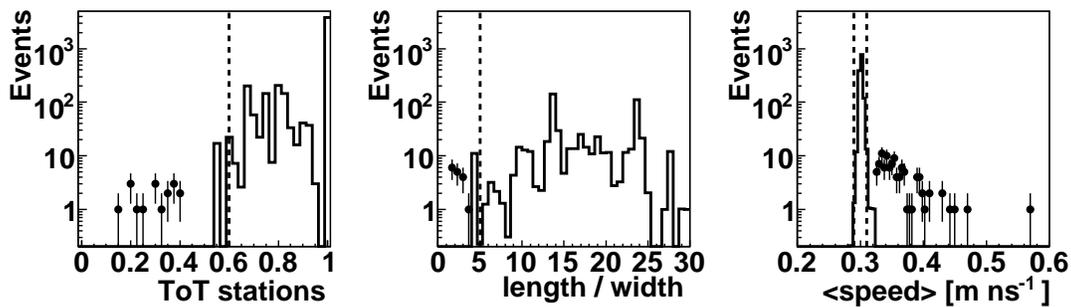}
  \caption{Distribution of discriminating variables for an incident  
  $E^{-2}$ $\nu_\tau$ flux (histogram), and for the real events (points). Events that survive all the selection cuts except the one for the shown variable are used. The vertical lines indicate the values to discriminate. Left panel: fraction of stations with a ToT signal;
  middle: length/width ; right: average of the ground speed between pairs of stations.}
  \label{fig:HillasCombined}
\end{figure}

The data have been searched for neutrino candidates over the analysed data 
period and there are no events fulfilling the selection cuts. The events failing to pass only one of them have a quite different distribution for the discriminant parameter than the one of the simulated $\tau$ showers (see figure~\ref{fig:HillasCombined}). In Table \ref{tab:events}, the number of events
surviving after successive cuts for real data as well as the efficiencies for simulated $\nu_\tau$s are
shown. The huge reduction of events after selecting very inclined showers (Elongated footprint and $Ground \ speed$) reaching the ground in early stages (Young showers) is expected for showers induced by protons or nuclei (see Section~\ref{signature}). The expected number of events surviving the combination of those cuts due to detector effects is also compatible with the observed discriminating power. Based on that, a limit for an injected spectrum $K \cdot \Phi(E)$ with a known shape
$\Phi(E)$ can be derived. The 90$\%$ confidence-level (CL), for 0 candidates and no background expected~\cite{FCStat}, on the value of $K$ is:

\begin{equation}
K_{90}=\frac{2.44}{\int{\Phi(E) \cdot Exp(E) dE}}
\end{equation}
where $Exp$ is interpolated from Table \ref{tab:uncertainties}.

\begin{table}[h]
\begin{center}
\fontsize{9}{10} \selectfont
\begin{tabular}{|c|c|c|}
\hline
Selection Requirement & MC Efficiency & Number of real events \\
\hline \hline
Initial sample & 1.00 & 3.97$\times$10$^{6}$\\
Young showers & 0.88 & 6.68$\times$10$^{5}$\\
Elongated footprint & 0.87 & 8.37$\times$10$^{3}$\\
$Ground \, speed \sim$ c & 0.84 & 0 \\
Contained footprint & 0.76 & 0 \\
\hline
\end{tabular}
\caption{\label{tab:events} Number of events passing the successive selection
  cuts. The Monte Carlo efficiency corresponds to identification
  efficiencies for $\nu$s of energy 1 EeV that
  trigger the SD detector. Data have been collected from January
  2004 until April 2008.}
\end{center}
\end{table}

\begin{table}[h]
\begin{center}
\fontsize{9}{10} \selectfont
\begin{tabular}{|c|cc|}
\hline
& \multicolumn{2}{|c|}{Exposure [$cm^2$ s sr] } \\
log($E_{\nu}$/eV) & Highest & Lowest \\
\hline \hline
17.0 & 1.81 $\times 10^{14}$ & 5.72 $\times 10^{12}$ \\
17.5 & 7.29 $\times 10^{15}$ & 1.93 $\times 10^{15}$ \\
18.0 & 2.44 $\times 10^{16}$ & 8.75 $\times 10^{15}$ \\
18.5 & 5.30 $\times 10^{16}$ & 1.98 $\times 10^{16}$ \\
19.0 & 7.17 $\times 10^{16}$ & 2.81 $\times 10^{16}$ \\
19.5 & 1.11 $\times 10^{17}$ & 3.41 $\times 10^{16}$ \\
20.0 & 1.18 $\times 10^{17}$ & 3.50 $\times 10^{16}$ \\
20.5 & 1.39 $\times 10^{17}$ & 3.41 $\times 10^{16}$ \\
\hline
\end{tabular}
\caption{\label{tab:uncertainties} Exposure of the SD of the Pierre Auger Observatory 
  from January 2004 until April 2008. In the columns labeled highest and lowest we give 
  the values of the exposure for the most optimistic and most pessimistic cases of the
  systematic uncertainties.}
\end{center}
\end{table}

For an injected diffuse flux of $\nu_{\tau}$ $dN/dE =
K \cdot E^{-2}$, the 90$\%$ CL limit is $K_{90}=6_{-3}^{+3} \times
10^{-8}$ GeV cm$^{-2}$ s$^{-1}$ sr$^{-1}$, where
the uncertainties come from the systematics discussed in section
\ref{systematic}. The bound is obtained for the 
energy range $2 \times 10^{17}$ - $2 \times10^{19}$ eV, with a systematic uncertainty of about 15$\%$, over which 90$\%$ of the events can be expected for an $E^{-2}$ flux. 

\begin{figure}[h]
   \centering
  \includegraphics[width=0.8\textwidth]{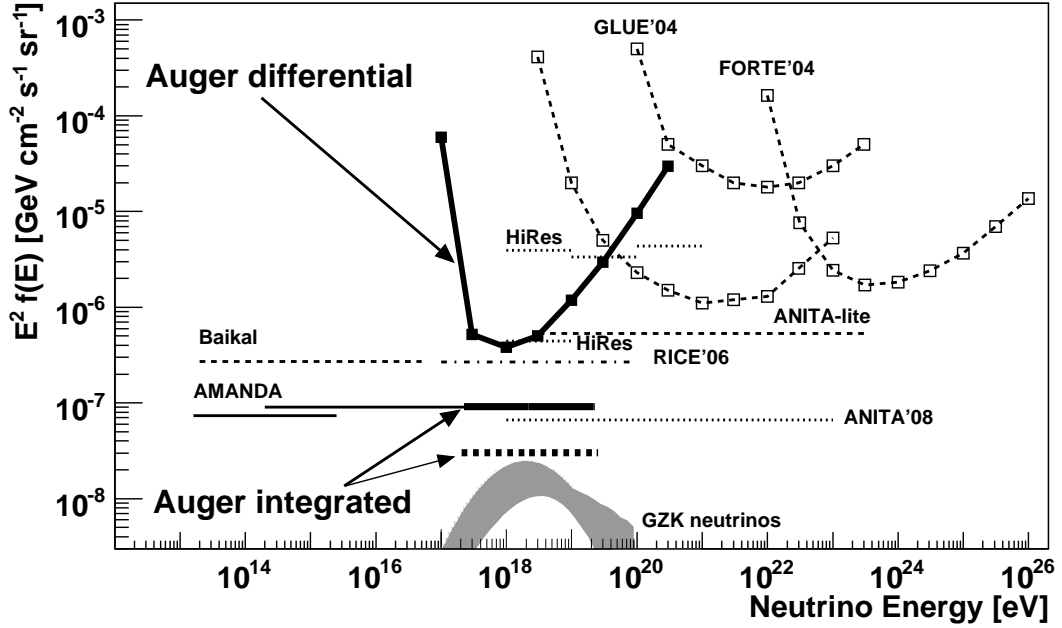}
  \caption{Limits at 90 $\%$ CL for {\it each flavor} of diffuse UHE 
    neutrino fluxes assuming a proportion of flavors of 1:1:1
    due to neutrino oscillations. The Auger limits are given using the most pessimistic case of the systematics (solid lines). For the integrated format, the limit that would be obtained in the most optimistic scenario of systematics is also shown (dashed line). See text for the references to the other experimental limits. The shaded area corresponds to the
    allowed region of expected GZK neutrino fluxes computed under
    different assumptions~\cite{Allard:2006mv,Protheroe99,Engel:2001hd,Anchordoqui:2007fi}, although predictions almost 1 order of magnitude lower and higher exist.}
  \label{fig:limits}
\end{figure}

In figure~\ref{fig:limits}, the limit for the {\it most pessimistic}
scenario of systematic uncertainties is shown. It improves by a factor
$\sim$ 3 in the most optimistic scenario (dotted line). Flux limits given by other
experiments are also shown (divided by 3 if they are limits to all flavours to be able to compare): 
AMANDA~\cite{Achterberg:2007nx,Ackermann:2005sb}, Baikal
~\cite{Aynutdinov:2005dq}, RICE~\cite{RICE05} (rescaled at 90 $\%$ CL),
HiRes~\cite{Martens:2007ff,Abbasi:2008hr}, ANITA-lite~\cite{Barwick:2005hn}, ANITA~\cite{Gorham:2008yk}, GLUE~\cite{Gorham:2003da} and FORTE~\cite{Lehtinen:2003xv}. For some of them the
limits are given for an $E^{-2}$ flux (integrated format), while for others the flux
limit is given as 2.3/Exposure$\cdot E_{\nu}$ (differential format). The limit from
the ANITA-lite experiment and the Pierre Auger Observatory are given in both formats for
comparison. The differential format demonstrates explicitly that the sensitivity of the Pierre Auger Observatory to Earth-skimming neutrinos peaks in a narrow energy range close to where the GZK neutrinos are expected.\par

The energy range of the $\nu_\tau$s explored with the Auger Observatory with this channel
is very well suited to search for the diffuse flux of $\nu$s that are produced by the GZK effect.\par

\section{Summary and Prospects}
\label{summary}

The dataset collected during the construction phase 
of the surface detector of the Pierre Auger
Observatory from January 2004 until April 2008, 
is used to present an upper limit to the diffuse flux of 
$\nu_{\tau}$. The Earth-skimming technique together with the
configuration of the surface detector gives the best sensitivity
currently available around a few EeV, which is the most relevant energy
to explore the predicted fluxes of GZK neutrinos. 
However in the worst case of systematic uncertainties, the limit presented here is
still higher by about one order of magnitude than GZK neutrino predictions. 
The Pierre Auger Observatory will keep taking data for about 20 years
over which the bound will improve by more than an order of magnitude
if no neutrino candidate is found.\par

\section*{Acknowledgments}

The successful installation and commissioning of the Pierre Auger Observatory
would not have been possible without the strong commitment and effort
from the technical and administrative staff in Malarg\"ue.

We are very grateful to the following agencies and organizations for financial support: 
Comisi\'on Nacional de Energ\'ia At\'omica, 
Fundaci\'on Antorchas,
Gobierno De La Provincia de Mendoza, 
Municipalidad de Malarg\"ue,
NDM Holdings and Valle Las Le\~nas, in gratitude for their continuing
cooperation over land access, Argentina; 
the Australian Research Council;
Conselho Nacional de Desenvolvimento Cient\'ifico e Tecnol\'ogico (CNPq),
Financiadora de Estudos e Projetos (FINEP),
Funda\c{c}\~ao de Amparo \`a Pesquisa do Estado de Rio de Janeiro (FAPERJ),
Funda\c{c}\~ao de Amparo \`a Pesquisa do Estado de S\~ao Paulo (FAPESP),
Minist\'erio de Ci\^{e}ncia e Tecnologia (MCT), Brazil;
AVCR AV0Z10100502 and AV0Z10100522,
GAAV KJB300100801 and KJB100100904,
MSMT-CR LA08016, LC527, 1M06002, and MSM0021620859, Czech Republic;
Centre de Calcul IN2P3/CNRS, 
Centre National de la Recherche Scientifique (CNRS),
Conseil R\'egional Ile-de-France,
D\'epartement  Physique Nucl\'eaire et Corpusculaire (PNC-IN2P3/CNRS),
D\'epartement Sciences de l'Univers (SDU-INSU/CNRS), France;
Bundesministerium f\"ur Bildung und Forschung (BMBF),
Deutsche Forschungsgemeinschaft (DFG),
Finanzministerium Baden-W\"urttemberg,
Helmholtz-Gemeinschaft Deutscher Forschungszentren (HGF),
Ministerium f\"ur Wissenschaft und Forschung, Nordrhein-Westfalen,
Ministerium f\"ur Wissenschaft, Forschung und Kunst, Baden-W\"urttemberg, Germany; 
Istituto Nazionale di Fisica Nucleare (INFN),
Ministero dell'Istruzione, dell'Universit\`a e della Ricerca (MIUR), Italy;
Consejo Nacional de Ciencia y Tecnolog\'ia (CONACYT), Mexico;
Ministerie van Onderwijs, Cultuur en Wetenschap,
Nederlandse Organisatie voor Wetenschappelijk Onderzoek (NWO),
Stichting voor Fundamenteel Onderzoek der Materie (FOM), Netherlands;
Ministry of Science and Higher Education,
Grant Nos. 1 P03 D 014 30, N202 090 31/0623, and PAP/218/2006, Poland;
Funda\c{c}\~ao para a Ci\^{e}ncia e a Tecnologia, Portugal;
Ministry for Higher Education, Science, and Technology,
Slovenian Research Agency, Slovenia;
Comunidad de Madrid, 
Consejer\'ia de Educaci\'on de la Comunidad de Castilla La Mancha, 
FEDER funds, 
Ministerio de Ciencia e Innovaci\'on,
Xunta de Galicia, Spain;
Science and Technology Facilities Council, United Kingdom;
Department of Energy, Contract No. DE-AC02-07CH11359,
National Science Foundation, Grant No. 0450696,
The Grainger Foundation USA; 
ALFA-EC / HELEN,
European Union 6th Framework Program,
Grant No. MEIF-CT-2005-025057, 
and UNESCO.




\end{document}